\documentclass[twocolumn,aps,pra,showpacs]{revtex4}

\usepackage{graphicx}

\begin{document}
\title{Proposed experiment for the quantum ``Guess My Number'' protocol}
\author{Ad\'{a}n Cabello}
\email{adan@us.es}
\author{Antonio J. L\'{o}pez-Tarrida}
\email{tarrida@us.es}
\affiliation{Departamento de F\'{\i}sica Aplicada II,
Universidad de Sevilla, 41012 Sevilla, Spain}
\date{\today}


\begin{abstract}
An experimental realization of the entanglement-assisted ``Guess
My Number'' protocol for the reduction of communication
complexity, introduced by Steane and van Dam, would require
producing and detecting three-qubit GHZ states with an efficiency
$\eta > 0.70$, which would require single photon detectors of
efficiency $\sigma > 0.89$. We propose a modification of the
protocol which can be translated into a real experiment using
present-day technology. In the proposed experiment, the quantum
reduction of the multiparty communication complexity would require
an efficiency $\eta > 0.05$, achievable with detectors of $\sigma
> 0.47$, for four parties, and $\eta > 0.17$ ($\sigma >
0.55$) for three parties.
\end{abstract}


\pacs{03.67.Hk,
02.50.Le,
03.65.Ud,
03.65.Ta}
\maketitle


One of the most impressive applications of quantum resources for
information processing is the reduction of the communication
complexity required for certain
computations~\cite{CB97,BCW98,BvHT99,Raz99,Sv00}. Let us suppose
that two or more separated parties need to compute a function of a
number of inputs distributed among them. Using the best classical
strategy, this would require a certain minimum amount of classical
communication to be transmitted between the parties. However, if
the parties initially shared some entangled states, then the
amount of classical communication required for the computation
would be a great deal smaller than if no entanglement were
present. The quantum advantage usually grows with the number of
parties involved~\cite{BCW98}. Entanglement-assisted reduction of
classical communication complexity has numerous potential
applications in computer networks, VLSI circuits, and data
structures~\cite{KS97}.

A particularly attractive, thought-provoking, and stimulating way
to show the quantum advantage was proposed by Steane and van Dam
as a method for always winning the television contest ``Guess My
Number'' (GMN)~\cite{Sv00}. A team of three contestants (Alice,
Bob, and Charlie), each of them isolated in a booth, is given an
integer number $n=n_A+n_B+n_C$ of apples (where $n_j=0$, $1/2$,
$1$, or $3/2$). One of the contestants must guess whether the
number is odd or even just by receiving one bit from the other two
contestants. The best classical strategy would allow the
contestants to win in 75\% of the cases. However, they can win in
100\% of the cases if they initially share three-qubit
Greenberger-Horne-Zeilinger (GHZ) states~\cite{GHZ89,GHSZ90}. The
same game can be played with four contestants and the quantum
versus classical advantage is the same: 100\% vs 75\%. Steane
and van Dam stressed that ``A laboratory demonstration of
entanglement-enhanced communication would be (\ldots) a landmark
in quantum physics and quantum information science''~\cite{Sv00}.
So far, however, the requirements for an experimental
implementation of the quantum GMN protocol have impeded further
progress. Some progress has been reported on simpler schemes of
quantum reduction of classical communication complexity. For
instance, Xue {\em et al.} presented an experiment on quantum
reduction of two-party communication complexity based on two-qubit
entanglement~\cite{XHZLG01}. Galv\~{a}o proposed a protocol
requiring only one qubit and a detection efficiency $\sigma
> 0.33$~\cite{Galvao02}. More recently, Brukner, \.{Z}ukowski, and
Zeilinger have introduced a quantum reduction of two-party
communication complexity based on the entanglement between two
qutrits~\cite{BZZ02}.

The main obstacle for an experimental realization of the quantum
GMN protocol is the high detection efficiency required. The
required setup would consist of a source of GHZ states, single
qubit operations, and single qubit detectors. If we define the
overall efficiency $\eta$ as the number of three-qubit (or four-qubit) joint
detections corresponding to GHZ states, divided by the number of
three-qubit (or four-qubit) systems emitted by the source, then,
assuming that when no joint detection occurs the probability of
winning the game is only $1/2$, the experimental probability of
winning the GMN game using GHZ states is
\begin{equation}
P_{\rm exp} (\eta) = \eta + (1-\eta) \frac{1}{2}.
\label{pexp}
\end{equation}
Therefore, the quantum advantage could be detected if an overall
efficiency $\eta > 0.50$ could be achieved. In the three-qubit
case, the experiment would require threefold coincidences between
detectors so that each individual detector should have an
efficiency $\sigma = 0.79$ (since $\sigma = \eta^{1/c}$, $c$ being
the number of qubits). Moreover, in order to obtain an
experimental quantum probability of winning the GMN game 10\%
higher than the best classical probability, we would need $\eta >
0.70$, which would require detectors of efficiency $\sigma >
0.89$.

Quantum optics provides the best way to produce qubits in a GHZ
state and distribute them to various spacetime regions. However,
the first experiments producing three-photon
polarization-entangled GHZ states~\cite{BPDWZ99,PBDWZ00} did not
satisfy the demands of the GMN protocol, because only a tiny
fraction of the ensemble of photon triplets was
detected~\cite{Sv00}. Further experiments producing four-photon
GHZ states~\cite{PDGWZ01} yield a fourfold coincidence with a
success probability $4$~times higher than that of previous
three-photon experiments. Moreover, recent
experiments~\cite{ZYCZZP03} report a fourfold coincidence rate
$2$~orders of magnitude brighter than in~\cite{PDGWZ01}. We shall
show that in the very near future this technology could allow an
experimental demonstration of a quantum reduction of a genuine
three or four-party communication complexity. In this paper, we
introduce a modified version of the quantum GMN protocol which is
experimentally feasible with current technology. We shall describe
a quantum reduction of three-party (four-party) communication
complexity in which the quantum advantage is clear, provided we
can produce three (four) qubits in a GHZ state and detect them all
separately with an overall efficiency $\eta > 0.17$, which would
require detectors of efficiency $\sigma > 0.55$ ($\eta > 0.05$ and
$\sigma > 0.47$, for four qubits). The main goal of this proposal
is to note that the absence of perfect sources and detectors does
not prevent us from performing an experimental demonstration of a
quantum reduction of a genuine multiparty communication
complexity and to stimulate experimental work along these lines.


The modified GMN game preserves all the essential features of the
original game, but includes rules that relax the detection
requirements to experimentally show the quantum advantage. The
modified GMN game features one referee (and a fourth contestant in
the $c=4$ version). We shall discuss in detail the four-party
version of the modified protocol; similar rules apply to the
three-party version. During the game, each of four contestants
(Alice, Bob, Charlie, and David) is isolated in a booth. Before
the game starts, they can take anything they want with them into
the booths, but once they are in, they will not be able to
communicate with each other or with anybody else, save for the
referee. Once they are in the booths, the referee distributes
among them a randomly chosen integer number $n$ of apples in four
portions, $n=n_A+n_B+n_C+n_D$, such that $n_j=0$, $1/2$, $1$, or
$3/2$. Then, the referee asks each and everyone whether or not
they are ready to play the game; if all contestants say yes, then
the referee asks Bob, Charlie, and David to give him a bit. Then,
the referee adds (modulo 2) the three bits, and hands the result
over to Alice. The team wins if Alice ascertains whether the total
number of distributed apples is even or odd. If any contestant
refuses to play the game, then the referee distributes a new
number $n'=n'_A+n'_B+n'_C+n'_D$ of apples and asks the four
contestants again whether or not they are ready to play the game,
etc. If the referee distributes $N$ rounds of apples, then the
contestants are forced to play the game for at least $r$ rounds
(hereafter referred as ``the played rounds''). The contestants
know $p=r/N$ before the game starts. In addition, the referee must
ensure that each of the $128$~possible variations of apples (see
Table~I) occurs with the same frequency in the played rounds.


\begin{table}[tbp]
\begin{center}
\begin{tabular}{ccc}
\hline \hline \multicolumn{1}{c}{$\;\{n_i,n_j,n_k,n_l\}\;$} &
$\;n_i+n_j+n_k+n_l\;$ & $\;$number of variations$\;$ \\ \hline
$\{0,0,0,0\}$ & $0$ & $1$ \\
$\{0,0,0,1\}$ & $1$ & $4$ \\
$\{0,0,1/2,1/2\}$ & $1$ & $6$ \\
$\{0,0,1/2,3/2\}$ & $2$ & $12$ \\
$\{0,0,1,1\}$ & $2$ & $6$ \\
$\{0,1/2,1/2,1\}$ & $2$ & $12$ \\
$\{1/2,1/2,1/2,1/2\}$ & $2$ & $1$ \\
$\{0,0,3/2,3/2\}$ & $3$ & $6$ \\
$\{0,1/2,1,3/2\}$ & $3$ & $24$ \\
$\{0,1,1,1\}$ & $3$ & $4$ \\
$\{1/2,1/2,1/2,3/2\}$ & $3$ & $4$ \\
$\{1/2,1/2,1,1\}$ & $3$ & $6$ \\
$\{0,1,3/2,3/2\}$ & $4$ & $12$ \\
$\{1/2,1/2,3/2,3/2\}$ & $4$ & $6$ \\
$\{1/2,1,1,3/2\}$ & $4$ & $12$ \\
$\{1,1,1,1\}$ & $4$ & $1$ \\
$\{1/2,3/2,3/2,3/2\}$ & $5$ & $4$ \\
$\{1,1,3/2,3/2\}$ & $5$ & $6$ \\
$\{3/2,3/2,3/2,3/2\}$ & $6$ & $1$ \\
\hline \hline
\end{tabular}
\end{center}
\noindent TABLE I. {\small The 19~integer combinations of $0$,
$1/2$, $1$, and $3/2$, and their corresponding $128$~variations.
In $64$ of them $n_i+n_j+n_k+n_l$ is an odd number while in the
other $64$ it is an even number.}
\end{table}

\begin{figure}
\centerline{\includegraphics[width=8.4cm]{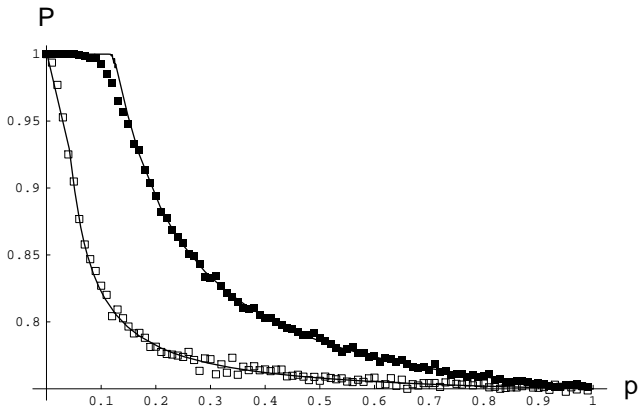}}
\caption{\label{Prw} Exact and numerical simulations of the
probability of the contestants winning the modified GMN game using
the best classical strategy, as a function of the minimum
percentage of rounds the referee forces them to play, for the
three-contestant game (black squares) and four-contestant game
(white squares). In the numerical simulations the referee
distributes $N = 100$ rounds. The exact probabilities are given by
Eq.~(\ref{Pn}). Interestingly, for $c=3$ contestants forced to
play in at least $p = 0.17$ of the rounds, the best classical
probability of winning is only $P_C = 0.92$. For $c=4$ contestants
forced to play in at least $p = 0.05$ of the rounds, the best
classical probability of winning is only $P_C = 0.90$.}
\end{figure}

In the modified GMN game, if the referee forces the contestants to
play in $r=p N$ of the $N$ rounds, the contestants can refuse to
play between the first and the $N-r$ round, but then they are
forced to play in the remaining $r$ rounds. If they decide to play
without being forced to do so then, every time they play, they
will postpone in one round the moment they have to play
compulsorily. The maximum classical probability of winning is
obtained by combining two strategies. The first one applies in the
rounds in which they play without being forced to do so, and can
be designed in a way such that the contestants know when they must
play and success is guaranteed when they do play (this happens, at
best, once in every $32$ rounds, on average, if $c=4$, and once in
every $8$ rounds, on average, if $c=3$). The second strategy
applies when they are forced to play. It could be any of the best
classical strategies of the original GMN game, giving a
probability of success of $3/4$ (for instance, each contestant
would give the referee a bit value $0$ if she/he had received
$n_j=0$ or $1/2$, or a bit value $1$ if she/he had received
$n_j=1$ or $3/2$). From all this follows that, for the modified
GMN game, the best classical strategies (of which there are
several) give the following maximum probability of winning for
$c=3$ or $c=4$ contestants being forced to play in at least $p$ of
the rounds,
\begin{eqnarray}
P_C(c,p) & = & \lim_{N \rightarrow \infty} \left[
\frac{(1-\mu)^{N-p N}}{4 p N}
\sum_{j=0}^{p N} (j+3p N) \mu^j
\right. \nonumber \\ & & \times \left(\begin{array}{c} N-p N+j-1
\\ j \end{array} \right) \nonumber \\ & & \left. + \sum_{j=p
N+1}^N (1-\mu)^{N-j} \mu^j \left(\begin{array}{c} N \\ j
\end{array} \right) \right],
\label{Pn}
\end{eqnarray}
where
\begin{equation}
\mu = \frac{8}{2^{2 c}}.
\end{equation}
For $c=3$ and $c=4$, this probability is represented as a function
of $p$ in Fig.~\ref{Prw}. In addition, Fig.~\ref{Prw} contains
numerical simulations of the probability that the team with $c=3$
and $c=4$ wins when using the best classical strategy for games of
$N=100$ rounds.


Note that, in both cases, if the referee forces the team to play
all the rounds, the probability of winning by using the best
classical strategy is $3/4$ while, if the referee forces them to
play in at least one of every 100~rounds, then the probability of
success using the best classical strategy is approximately $1$.


Let us now see what the probabilities of winning are when using
the best entanglement-assisted strategy. The contestants will
always win if they use the following method:

(1) Each contestant carries a qubit belonging to a four-qubit
system initially prepared in the GHZ state
\begin{equation}
|{\rm GHZ}\rangle = {1 \over \sqrt{2}} (|\bar{0} \bar{0} \bar{0}
\bar{0} \rangle+|\bar{1} \bar{1} \bar{1} \bar{1}\rangle),
\label{GHZ03}
\end{equation}
where $|\bar{0} \bar{0} \bar{0} \bar{0}\rangle = |\bar{0}\rangle
\otimes |\bar{0}\rangle \otimes |\bar{0}\rangle \otimes
|\bar{0}\rangle$, where $|\bar{0}\rangle = (1 / \sqrt{2})
(|0\rangle + |1\rangle)$ and $|\bar{1}\rangle = (1 / \sqrt{2})
(|0\rangle -|1\rangle)$.

(2) Each contestant $j$ applies to her/his qubit the rotation
\begin{equation}
R(n_j)=|\bar 0\rangle\langle \bar 0|+e^{i n_j \pi}|\bar
1\rangle\langle \bar 1|, \label{rotj}
\end{equation}
where $n_j$ is her/his number of apples.

(3) Then, each contestant measures her/his qubit in the
computational basis $\{|0\rangle, |1\rangle\}$.

(4) If, due to the inefficiency of the detectors, a contestant
does not obtain a result, then she/he will tell the referee that
she/he will not play the game, and the referee will therefore
abort that round. Note that, in the aborted rounds, Alice does not
receive any bits from the referee. If all contestants consent to
play that round, then Bob, Charlie, and David will give their
outcomes to the referee, who will add them up, and give the result
to Alice.

In this case Alice can give the correct answer with probability
$1$ because state~(\ref{GHZ03}) has the following property: for
any $n_A+n_B+n_C+n_D$ integer (where $n_j=0$, $1/2$, $1$, or
$3/2$),
\begin{eqnarray}
\lefteqn{R(n_A) \otimes R(n_B) \otimes R(n_C) \otimes R(n_D)
|{\rm GHZ}\rangle} \nonumber \\
& & = \left\{\begin{array}{ll}
|{\rm GHZ}\rangle & \mbox{if $n_A+n_B+n_C+n_D$ is even,} \\
|{\rm GHZ}^\perp\rangle & \mbox{if $n_A+n_B+n_C+n_D$ is odd,}
\end{array} \right.
\end{eqnarray}
where $|{\rm GHZ}\rangle$ and $|{\rm GHZ}^\perp\rangle$ can be
reliably distinguished by local measurements in the computational
basis:
\begin{eqnarray}
|{\rm GHZ}\rangle & = & {1 \over 2 \sqrt{2}} (|0000\rangle
+|0011\rangle+|0101\rangle +|0110\rangle\nonumber \\
& & +|1001\rangle+|1010\rangle+|1100\rangle+|1111\rangle),
\label{GHZ03comp} \\
|{\rm GHZ}^\perp\rangle & = & {1 \over 2 \sqrt{2}}
(|0001\rangle+|0010\rangle+|0100\rangle +|0111\rangle\nonumber \\
& & +|1000\rangle+|1011\rangle+|1101\rangle+|1110\rangle).
\label{GHZ03perpcomp}
\end{eqnarray}

Assuming that, when all four contestants obtain a result, this
corresponds to a GHZ state (i.e., assuming that any error in the
preparation is negligible), then having an experimental efficiency
$\eta$ allows the team to play the modified GMN game with
$p=\eta$. Now let us go back to the probabilities illustrated in
Fig.~\ref{Prw}. In the first place we shall compare the
experimental requirements for the original GMN game with three
qubits to those of the modified protocol. The most important point
is that, while in the original GMN protocol the difference between
the quantum and classical probabilities of winning could be
detected only if the experimental setup has an overall efficiency
$\eta > 0.50$ (that is, a single qubit detection efficiency
$\sigma = 0.79$), in the modified protocol the difference between
the quantum and classical probabilities {\em can be detected for
almost any efficiency}. Moreover, as seen above, to obtain an
experimental quantum probability of winning 10\% higher than the
best classical probability in the original GMN protocol, the setup
would need to have $\eta > 0.70$ (that is, a single qubit
detection efficiency $\sigma > 0.89$). However, to obtain a
difference between the quantum and classical probabilities of
winning higher than $7.7$\% in the modified protocol, the setup
would only require $\eta > 0.17$ (that is, detectors of efficiency
$\sigma > 0.55$). On the other hand, since sources of four-photon
GHZ states (\ref{GHZ03}) are currently
available~\cite{PDGWZ01,ZYCZZP03}, then it is interesting to note
that, for $c=4$ contestants and an experimental setup with an
overall efficiency $\eta > 0.05$ (that is, with detectors of
efficiency $\sigma > 0.47$), it would be possible to obtain a
difference between the quantum and classical probabilities higher
than $9.7$\%. Photodetectors of $\sigma > 0.47$ are currently
available. Other examples for different values of $\sigma$ can be
found in Table~II. An interesting advantage of all these
experiments is that the expected quantum probabilities are 1,
which implies that the error of the experimental results, given by
the standard deviation $\sqrt{P (1-P)}/r$, where $r$ is the number
of coincidences (i.e., played rounds), should be very low.


\begin{table}[tbp]
\begin{center}
\begin{tabular}{cccc}
\hline \hline $\;\;\;c\;\;\;$ & $\;\;\;P_Q-P_C\;\;\;$ &
$\;\;\;\;\;\;\eta\;\;\;\;\;\;$ & $\;\;\;\;\;\;\sigma\;\;\;\;\;\;$
\\ \hline
$3$ & $0.250$ & $1$ & $1$ \\
$4$ & $0.250$ & $1$ & $1$ \\
$3$ & $>0.214$ & $>0.50$ & $>0.79$ \\
$4$ & $>0.218$ & $>0.20$ & $>0.67$ \\
$3$ & $>0.107$ & $>0.20$ & $>0.58$ \\
$4$ & $>0.177$ & $>0.10$ & $>0.56$ \\
$3$ & $>0.077$ & $>0.17$ & $>0.55$ \\
$4$ & $>0.097$ & $>0.05$ & $>0.47$ \\
\hline \hline
\end{tabular}
\end{center}
\noindent TABLE II. {\small Examples of single photon detection
efficiency requirements for the modified GMN protocol. $c$ is the
number of parties, $P_Q-P_C$ is the difference between the quantum
and classical probabilities of winning, $\eta$ is the number of
joint detections divided by the number of systems emitted by the
source, and $\sigma$ is the corresponding single photon detection
efficiency.}
\end{table}


The proposed experiment would consist of a source emitting three
(or four) polarization-entangled photons in a GHZ state generated
in a parametric-down conversion process~\cite{PDGWZ01,ZYCZZP03},
coupled into three (four) single mode optical fibers which
distribute the photons to different regions, where each photon
suffers a randomly chosen rotation of the type~(\ref{rotj}), and a
linear polarization measurement (typically the horizontal and
vertical states represent the computational basis). If all photons
are detected, then two (three) of the contestants send their
result to the referee who adds them up and sends the result to the
third (fourth) contestant, who adds it to her result and gives the
answer.

To sum up, while testing the advantage of the original quantum GMN
protocol involving three parties would require detectors of an
efficiency {\em at least} $\sigma > 0.79$ (or $\sigma > 0.89$ to
obtain an experimental quantum probability of winning 10\% higher
than the best classical probability), we have introduced a
modified quantum GMN protocol involving three or four parties and
preserving all the essential features of the original one, but
with the remarkable property that the quantum vs classical
advantage is detectable for {\em any} $\sigma$. To be specific,
$\sigma > 0.55$ would allow us to obtain an experimental quantum
probability of winning at least $7.7$\% higher than the best
classical probability in the three-party case, and $\sigma
> 0.47$ would allow us to obtain an experimental quantum
probability of winning at least $9.7$\% higher than the best
classical probability in the four-party case. Our hope is that
this proposal will stimulate experimental work to detect the
quantum reduction of a genuine multiparty communication complexity.\\

The authors thank M.~Bourennane, E.~F.~Galv\~{a}o, C.~Serra, and
H.~Weinfurter for useful discussions and comments. This work was
supported by the Spanish Ministerio de Ciencia y Tecnolog\'{\i}a
Project~BFM2002-02815 and the Junta de Andaluc\'{\i}a
Project~FQM-239.




\begin{thebibliography}{99}

\bibitem{CB97}
R. Cleve and H. Buhrman,
Phys. Rev. A {\bf 56}, 1201 (1997).

\bibitem{BCW98}
H. Buhrman, R. Cleve, and A. Wigderson,
in {\em Proceedings of the 30th Annual ACM Symposium on the Theory
of Computing}
(ACM Press, New York, 1998), p. 63.

\bibitem{BvHT99}
H. Buhrman, W. van Dam, P. H\o{}yer, and A. Tapp,
Phys. Rev. A {\bf 60}, 2737 (1999).

\bibitem{Raz99}
R. Raz,
in {\em Proceedings of the 31st Annual ACM Symposium on the Theory of Computing}
(ACM Press, New York, 1999), p. 358.

\bibitem{Sv00}
A.M. Steane and W. van Dam,
Phys. Today {\bf 53} (2), 35 (2000).

\bibitem{KS97}
E. Kushilevitz and N. Nisan,
{\em Communication Complexity}
(Cambridge University Press, Cambridge, England, 1997).


\bibitem{GHZ89}
D.M.~Greenberger, M.A.~Horne, and A.~Zeilinger,
in {\em Bell's Theorem, Quantum Theory, and Conceptions of the
Universe}, edited by M. Kafatos
(Kluwer Academic, Dordrecht, Holland, 1989), p. 69.

\bibitem{GHSZ90}
D.M.~Greenberger, M.A.~Horne, A.~Shimony, and A.~Zeilinger,
Am. J. Phys. {\bf 58}, 1131 (1990).


\bibitem{XHZLG01}
P. Xue, Y.-F. Huang, Y.-S. Zhang, C.-F. Li, and G.-C. Guo,
Phys. Rev. A {\bf 64}, 032304 (2001).

\bibitem{Galvao02}
E.F. Galv\~{a}o,
Phys. Rev. A {\bf 65}, 012318 (2002).

\bibitem{BZZ02}
\v{C}. Brukner, M. \.{Z}ukowski, and A. Zeilinger,
Phys. Rev. Lett. {\bf 89}, 197901 (2002).


\bibitem{BPDWZ99}
D. Bouwmeester, J.-W. Pan, M. Daniell, H. Weinfurter, and A.
Zeilinger,
Phys. Rev. Lett. {\bf 82}, 1345 (1999).

\bibitem{PBDWZ00}
J.-W. Pan, D. Bouwmeester, M. Daniell,
H. Weinfurter, and A. Zeilinger,
Nature (London) {\bf 403}, 515 (2000).

\bibitem{PDGWZ01}
J.-W. Pan, M. Daniell, S. Gasparoni, G. Weihs, and A. Zeilinger,
Phys. Rev. Lett. {\bf 86}, 4435 (2001).

\bibitem{ZYCZZP03}
Z. Zhao, T. Yang, Y.-A. Chen, A.-N. Zhang, M. \.{Z}ukowski, and
J.-W. Pan,
Phys. Rev. Lett. {\bf 91}, 180401 (2003).


\end{thebibliography}
\end{document}